# Branching Visual Inquiry in a Generative Visual Browser: A Matched Walkthrough Study

Hanjing Shi, Kimberly Wang, and Dominic DiFranzo

Lehigh University

## Abstract

Generative visual browsers combine AI-generated visual pages with reader-selected branches rather than a fixed sequence. We analyzed eight author self-walkthroughs of Flipbook, organized as four pairs with identical starting prompts about sleep, vaccines, flight, and Pluto. Three paired routes differed at the first selection; the sleep routes separated after a shared repair-oriented opening. A route could remain locally coherent while narrowing to one mechanism or leaving part of its starting question unanswered. Some diagrams lacked sources. The analysis motivates three design directions: persistent path histories, prompts that reconnect a route to its starting question, and source or uncertainty cues for generated claims. It characterizes how one generative visual browser structured eight documented routes.

## 1. Objectives or Purposes

Generative AI is often encountered through answer-generation, summarization, or tutoring interfaces; multimedia learning identifies conditions for complementary visual and verbal representations (Kasneci et al., 2023; Mayer, 2009). The public web-based Flipbook browser examined here uses a reader-directed, hypertext-like interaction. A reader begins with a natural-language prompt, receives a generated visual page and suggested next topics, then selects an adjacent topic to continue (Flipbook, n.d.). Each selection produces another page and adds a breadcrumb. Here, a "branch" denotes one selected next topic, a "route" the ordered sequence of selections and pages, and a "breadcrumb" its visible record.

The format differs from a fixed e-book or pre-authored hypertext document (DeStefano & LeFevre, 2007). A route is co-produced: the system supplies pages and branch options, while a reader selects among them. This enables exploratory reading but creates an analytic problem: a sequence of related pages can appear coherent while leaving the starting question partly answered. The analysis therefore asks: (1) How did the two authors' routes converge or diverge after the same starting prompt, and what selection rationales did their memos document? (2) When did local page-to-page coherence coexist with an incomplete or misaligned answer to the starting prompt? (3) Which interface requirements follow from the documented pathways and breakdowns?

The corpus contains four matched-start pairs on sleep, vaccines, flight, and Pluto, each a question with more than one plausible explanation. Matching prompts supports comparison of shared questions and the routes that subsequently unfold without treating either route as the topic's correct structure. The study contributes an artifact-centered account in which a browsing session is an ordered record of a prompt, a selected branch, a generated page, and a breadcrumb. That record distinguishes page-to-page coherence from a route's cumulative answer to its initial question and identifies generated claims without visible support.

## 2. Perspective(s) or Theoretical Framework

Cognitive flexibility theory provides the main frame. It describes advanced understanding as the ability to reorganize knowledge across cases and perspectives rather than follow one fixed sequence (Spiro et al., 1988). A branching interface can make alternative routes through an ill-structured topic available, but that possibility matters only when a reader can see how successive pages relate.

Hypertext research identifies the difficulty in that condition. Links can support nonlinear inquiry, but choosing among them adds cognitive demands when readers must infer the organizing structure (DeStefano & LeFevre, 2007). Visual representations can clarify a mechanism when image and text provide complementary information (Mayer, 2009); they can also make an incomplete account appear settled. A route must therefore be examined as more than a sequence of topically related pages.

Self-regulated learning specifies what such exploration requires: monitoring the goal, evaluating the current step, and deciding whether to continue or redirect (Azevedo & Cromley, 2004; Zimmerman, 2002). Generated content also requires verification because fluent explanations may be incomplete or incorrect (Kasneci et al., 2023). These perspectives direct the analysis to branch divergence, local coherence, global alignment, and visible support for claims. Local coherence concerns the relation between consecutive pages; global alignment concerns whether the accumulated route still addresses the prompt.

Together, these perspectives connect the available branches with the work of maintaining a question across successive pages. Cognitive flexibility directs attention to changes in explanatory perspective; hypertext research directs attention to how a route is organized. Self-regulated learning makes the initial question a reference for monitoring that route, while multimedia learning distinguishes the presentation of a mechanism from its integration into a larger explanation. These connections organize our reading of the records: what a selection adds locally, how the accumulated explanation relates to the prompt, and which claims have visible attribution.

## 3. Methods, Techniques, or Modes of Inquiry

To trace those route-level conditions in an operating browser, we adapted the walkthrough method, which uses systematic navigation to document an interface's functions and implied uses (Light et al., 2018). In July 2026, two author-analysts independently explored the same four starting prompts: the biological necessity of sleep, how vaccines work, the four forces of flight, and why Pluto is a dwarf planet. The resulting eight routes form four matched-start pairs.

The adaptation focuses on everyday navigation through generated pages. The four pairs were assembled around topics for which both authors had a documented route. The broader project also included stock-market and Wi-Fi walkthroughs by Author 2; those records are outside this shared-prompt comparison. Within each pair, matching refers to the wording of the initial prompt. The route is then an unfolding sequence in which the system presents content and possible next topics and the author selects a continuation.

The common prompt, rather than path length, supplied the basis for comparison. Author 2 recorded six selections for every prompt. Author 1 recorded five selections for sleep, six for vaccines, four for flight, and five for Pluto, yielding 44 transitions. Letting each route end at its recorded endpoint preserved actual navigation, but route length is not treated as an outcome.

The project brief for structured walkthroughs invited approximately six to ten selections, allowing an endpoint when a route became coherent or clearly broke down. The present comparison retains each documented endpoint. All eight routes include at least four selections, so the differences at the first selection precede their unequal endings. The amount of later specialization remains dependent on the depth available in each record.

Navigation records established what each author selected; memos supplied the authors' accounts of why. Both authors wrote one reflective memo per walkthrough. Author 2 recorded a structured log and memo while exploring, whereas Author 1 wrote retrospective analytic memos after reviewing the screenshot record. Screenshots and breadcrumbs, not Author 1's retrospective memo, established Author 1's selections; the memo did not add missing nodes or clicks. Both records were reduced to common fields: node order, coded domain, relation between adjacent pages, contribution to the starting question, reported starting knowledge and selection rationale, and claims requiring verification.

Author 2 used a repeated entry format: the starting prompt, topic interest, and prior knowledge, followed by the current node, selected branch, selection rationale, screenshot, reported new content, domain, and relation to the preceding node. Post-walkthrough knowledge maps arranged the nodes in sequence, added relationship labels, and identified the strongest and weakest transitions. For Author 1, the screenshot sequence supplied the route reconstruction and the later memo supplied a reflective account. Bringing these records together preserved both the navigational trace and the author account associated with it.

The two authors then jointly examined every transition and marked convergence, divergence, repetition, narrowing, global drift, and claims without visible sources. A transition was convergent when both routes selected the same mechanism or explanatory level, even if page titles differed. A route was globally misaligned when its endpoint remained topically related but no longer helped answer the prompt. Screenshots were revisited whenever a title alone did not establish the relation. This placed local coherence at the transition level, global alignment at the route level, and source support at the page level.

The joint reading moved between adjacent pages and the full route retained with its prompt. Relationship descriptions, such as a component leading to a mechanism or a process leading to its energy requirement, explained local connections. Comparing those connections across a pair identified shared mechanisms and changes in explanatory scale. Reading the full sequence against its prompt then distinguished continued relevance from coverage of only one part of the question. Appendix A makes these analytic distinctions explicit. Table 1 connects representative transitions with route-level judgments and their evidence locations.

This comparison has a defined scope. Generated starting pages and branch menus varied across sessions, so the shared prompt was a common conceptual entry, not a set of identical interface options. The authors analyzed only their own navigation records; the dataset contains no participant, demographic, assessment, or third-party data, and institutional review was not required. The comparisons therefore describe observed route differences without assigning them independently to author choice or system generation.

## 4. Data Sources, Evidence, Objects, or Materials

These procedures produced 52 screenshots: 24 from Author 1 and 28 from Author 2. Together, they preserve four initial prompts, 44 selected transitions, generated pages, and breadcrumb histories. Eight reflective memos accompany the screenshots: four contemporaneous records from Author 2 and four retrospective accounts from Author 1. Table 1 relates local connections to the scope of each route. Figure 1 places the two vaccine examples together, and Figure 2 presents all eight recorded paths at aligned selection depths. A screenshot with its breadcrumb, rather than a topic label alone, documented each interface state.

Flipbook is the object of analysis (Flipbook, n.d.). The screenshots and breadcrumbs establish what the interface presented, not whether a scientific statement was accurate. For discrete verification questions, including Pluto's classification criterion, we consulted external domain sources.

## 5. Results and/or Substantiated Conclusions

The common prompts did not produce common explanatory routes. Three pairs diverged at the first selection: vaccines split into antigen presentation and antibody diversification, flight into thrust and lift, and Pluto into roundness and orbital geometry. The sleep routes shared restoration and DNA repair before separating. The analyses below follow what those early differences meant for the scope, alignment, and evidentiary support of the later explanations.

The decisive difference was usually explanatory scale rather than discipline. Most steps remained in neighboring scientific fields, yet the paired routes selected different objects and mechanisms. In the vaccine, flight, and Pluto pairs, the first branch set the object that later selections deepened: antigen presentation or antibody diversification; propulsion or lift; material physics or orbital geometry. A final page from either branch could appear relevant in isolation. Read with its starting prompt and breadcrumb, it often supplied only part of the answer.

The memos place those divergences in relation to different questions and starting knowledge. In sleep, Author 1 knew that sleep matters for the brain but not how DNA repair relates to sleep, and followed repair toward neuronal restoration. Author 2 followed repair toward its timing, a specific enzyme, and an energy source. In vaccines, Author 1 extended partial knowledge of T cells through MHC class II and peptide loading, whereas Author 2 pursued antibody specificity and diversification. In flight, Author 1 described the four-force model and Newton's third law as familiar but formula-based and used diagrams to make thrust more concrete; Author 2 pursued lift through pressure distribution and stall. Pluto was unfamiliar to Author 1, while Author 2 followed orbital comparison.

The click logs make this movement concrete. At the antibody-specificity selection, Author 2 wrote, "I wanted to understand why antibodies bind only to specific pathogens" (vaccines, click 2). At the ATP selection, the rationale was, "I wanted to understand where the energy comes from for the DNA repair process" (sleep, click 6). Each rationale followed a question raised by the preceding page, making the next step intelligible within the developing subtopic.

The sleep pair shows that divergence can emerge after a shared opening. After DNA repair, Author 1 followed PARP1 toward neuronal restoration and synaptic plasticity. Author 2 moved through circadian regulation, OGG1, base-excision repair, and ATP. One route widened toward neural consequences; the other narrowed toward biochemical machinery.

The vaccine pair shows a different problem: complementary mechanisms can still leave the original question partly answered. Author 1 followed dendritic-cell processing, MHC class II, the peptide-binding groove, HLA-DM, CLIP, and the invariant chain (Figure 1A). Author 2 followed antibodies, specificity, the variable region, V(D)J recombination, junctional diversity, and exonuclease activity (Figure 1B). One route addressed antigen presentation to T cells and the other antibody diversity, but neither returned to how vaccines work.

The flight routes make the same loss of synthesis visible in a system model. From the four-force overview, Author 1 traced thrust through Newton's third law, momentum transfer, and propeller-force vectors. Author 2 traced lift through pressure differences, pressure distribution, stagnation, flow separation, and stall. Neither route reintegrated thrust, lift, drag, and weight into the starting model.

The Pluto pair most clearly separates local coherence from global alignment. Author 1 followed hydrostatic equilibrium, self-gravity, mass thresholds, and material strength. Author 2 followed elliptical orbit, inclination, planetary comparison, Mercury, and solar gravity. Every step remained related to Pluto, yet neither selected branch sequence developed the decisive criterion that Pluto has not cleared its orbital neighborhood (NASA, n.d.). Author 1's initial page already displayed this criterion; the subsequent selections followed roundness and material strength instead.

Across all four pairs, breadcrumbs recorded order but not the explanatory relation between pages or the part of the starting question that remained unanswered. Local coherence and global alignment could therefore come apart: a route could be easy to follow while still leaving part of its starting question unanswered.

The post-walkthrough maps provide a useful second view of this distinction. Author 2 described the Pluto route as coherent while identifying the move from Mercury to solar gravity as its weakest transition: "The starting prompt of Pluto becomes less direct." The map thus retained both a positive judgment of page-to-page continuity and a concern about the route's relation to its starting question. Specialization itself can serve an exploratory purpose; the relevant issue is how the selected explanation is connected back to the question being pursued.

A separate evidentiary gap concerned the pages themselves. Specialized biological and aerodynamic diagrams did not show sources for their claims. The vaccine screenshots illustrate the issue: both pages are visually specific and consistent with their local breadcrumbs, but neither identifies the evidence behind its claims. The results therefore point to two distinct supports: reconnecting a route to its starting question and making the basis of specialist claims visible.

## 6. Scientific or Scholarly Significance of the Study or Work

A matched-start walkthrough retains the shared prompt with screenshots, branch sequences, and breadcrumbs, making routes comparable even as each click changes later pages and menus. It makes the earlier selections that frame a final page available for analysis and reveals plausible, non-equivalent routes from one question without treating either as a benchmark.

The study also distinguishes four analytic levels a polished browser can otherwise collapse: page legibility, transition coherence, route alignment, and claim provenance. The Pluto routes show that topical pages and a coherent sequence can still leave a classification criterion unexplored in subsequent branches. The vaccine and flight routes show that an interface can deepen one mechanism without reconnecting it to the starting model. This distinction identifies where support is present and where it remains missing.

The observed breakdowns yield design implications. Persistent path histories would make a route inspectable. Keeping the prompt beside the breadcrumb would make relevance visible at choice, and a prompt after several selections could ask what the route adds or whether another branch is needed. Generated claims should carry sources, provenance, or uncertainty cues.

These directions can be expressed as concrete interface proposals. A path panel could retain visited pages with return links, alongside the existing breadcrumb. A small prompt panel could keep the starting question visible and offer an optional checkpoint asking what the current branch explains and what remains to explore. A claim-level source link could connect a generated statement to its supporting material, with an uncertainty label when support is unavailable. These proposals translate the route observations into features for future design testing; their effects on browsing and learning remain questions for that work.

The contribution is bounded by one platform, two author-analysts, four prompts, unequal route lengths, and branch menus that varied across runs. It maps observed route differences and interface output rather than isolating author preference, comparing Flipbook with linear text, or assessing learning effects. Participant research can next examine whether the proposed supports change comprehension, recall, or transfer.

**Table 1. Route evidence at the transition and original-question levels.**

S = sleep; V = vaccines; F = flight; P = Pluto. The digit identifies the author. A locator such as V1.4 denotes Author 1's vaccine page after selection 4; step 0 is the entry page. Arrows describe consecutive recorded pages.

| Route | Local connection and evidence | Relation to the original question |
|---|---|---|
| Sleep<br>S1<br>5 steps | DNA repair → PARP1 (2→3): a repair process leads to a named molecular component. Neuronal restoration → synaptic plasticity (4→5) extends the account toward neural function. | Develops repair and neural consequences of sleep. The final extension introduces a broader set of claims without integrating the other functions of sleep. Evidence: S1.1–5. |
| Sleep<br>S2<br>6 steps | OGG1 → base-excision repair (4→5): an enzyme leads to its pathway. Base-excision repair → ATP (5→6) follows that pathway to its stated energy requirement. | Develops the timing and machinery of DNA repair. The selected sequence remains within this restorative function rather than integrating the broader necessity of sleep. Evidence: S2.1–6. |
| Vaccines<br>V1<br>6 steps | Peptide-binding groove → HLA-DM (3→4): structure leads to peptide-loading regulation. CLIP → invariant chain (5→6) returns to the precursor of a component already encountered. | Develops antigen processing and presentation to T cells. Later steps specialize in peptide-loading machinery without synthesizing the broader protective response following vaccination. Evidence: V1.1–6; Figure 1A shows V1.4. |
| Vaccines<br>V2<br>6 steps | Antigen specificity → variable region (2→3): binding specificity leads to its structural site. V(D)J recombination → junctional diversity (4→5) adds a source of variation. | The antibody entry connects vaccination to protection; later steps develop specificity and genetic diversity. The route does not reconnect those details into an account of vaccine-induced protection. Evidence: V2.1–6; Figure 1B shows V2.3. |

The examples identify relations displayed in the records. They do not establish the scientific accuracy of the generated explanations.

**Table 1 (continued). Flight and Pluto routes.**

| Route | Local connection and evidence | Relation to the original question |
|---|---|---|
| Flight<br>F1<br>4 steps | Newton's third law → momentum transfer (2→3): a governing principle leads to a propulsion mechanism. Momentum transfer → propeller vectors (3→4) adds a force representation. | Selects thrust from the initial four-force model and develops propulsion. The subsequent sequence does not reintegrate thrust with lift, drag, and weight. Evidence: F1.0–4. |
| Flight<br>F2<br>6 steps | Pressure differential → pressure distribution (2→3) maps the selected mechanism across the wing. Flow separation → stall (5→6) follows a process to its stated consequence. | Selects lift and develops pressure and airflow conditions. The subsequent sequence does not reintegrate this explanation with the other three forces. Evidence: F2.0–6. |
| Pluto<br>P1<br>5 steps | Hydrostatic equilibrium → gravity versus rigidity (1→2) explains roundness through competing forces. Rock strength → ice versus rock (4→5) adds a material comparison. | Develops roundness as one classification criterion. The entry page already displays failure to clear the orbital neighborhood, but the selected branches do not develop that criterion. Evidence: P1.0–5. |
| Pluto<br>P2<br>6 steps | Orbital inclination → Earth comparison (2→3) supplies a reference for comparison. Mercury → solar gravity (5→6) extends the route toward general orbital forces. | Develops orbital properties and planetary comparisons without explaining the orbital-clearing criterion. The map identifies the final transition as less directly connected to Pluto. Evidence: P2.1–6 and its post-walkthrough map. |

Note. Partial coverage is distinguished from loss of relevance. Selection depth is a record locator, not a measure of learning or explanation quality. The source key in Table A2 links route identifiers to the underlying records.

## Figure 1. Two vaccine branches from the same starting prompt.

### A Author 1: peptide loading (V1.4)

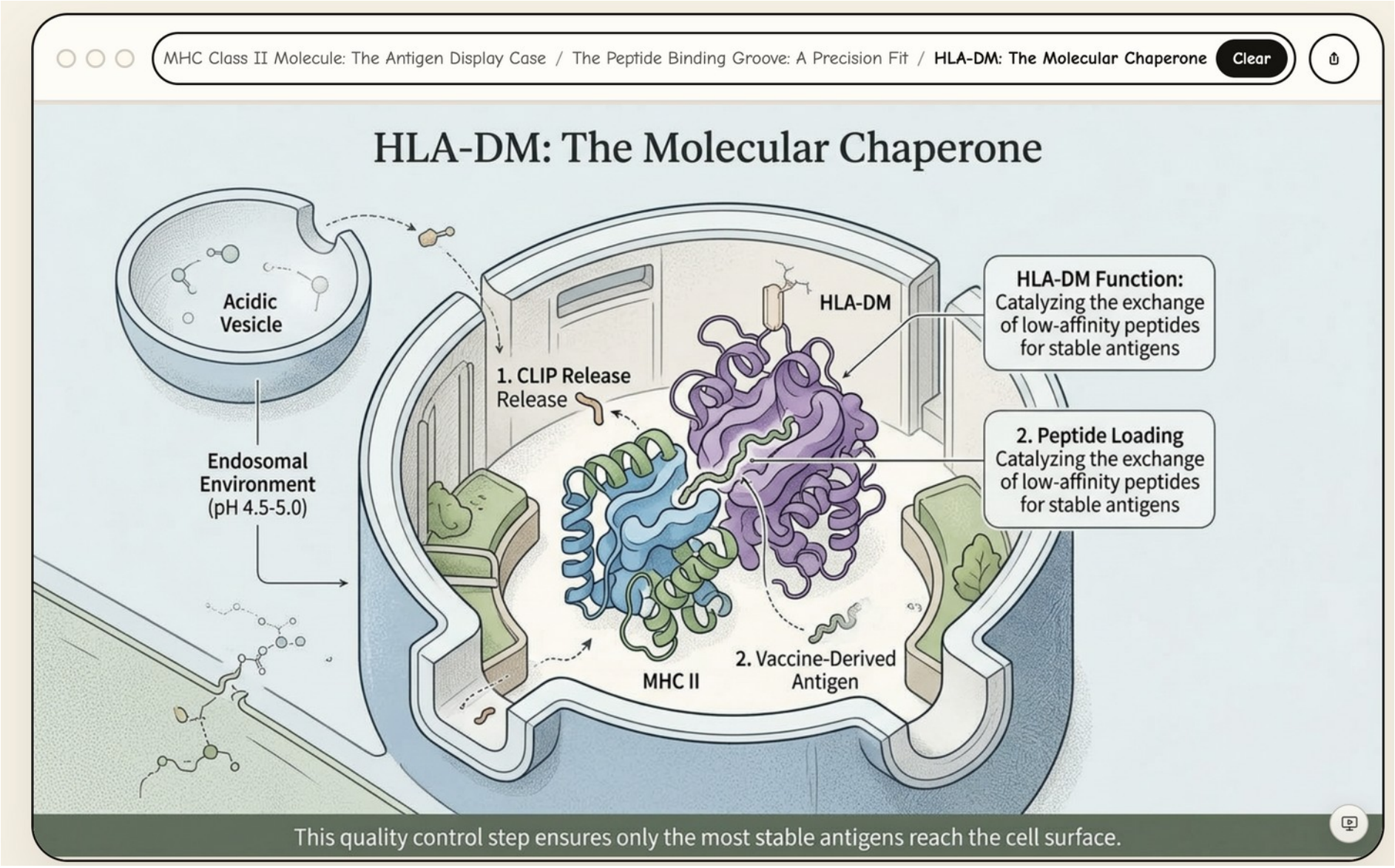

**Breadcrumb**
MHC class II → peptide-binding groove → HLA-DM.

**Displayed focus**
The diagram develops the exchange and loading of peptides.

**Route context**
Antigen presentation is one part of the broader vaccine response.

### B Author 2: antibody structure (V2.3)

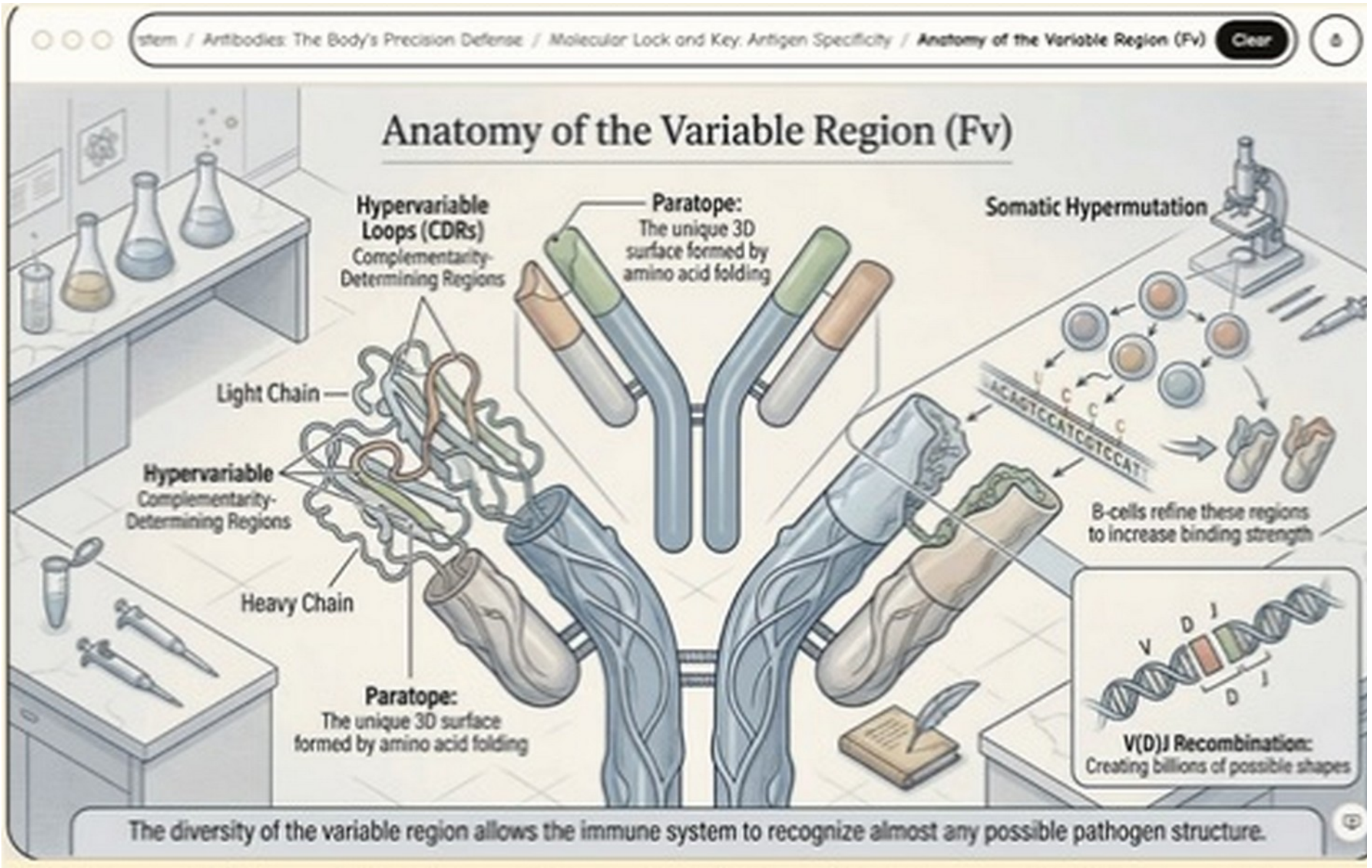

**Breadcrumb**
Antibodies → antigen specificity → variable region.

**Displayed focus**
The diagram develops the structural basis of antigen binding.

**Route context**
The subsequent route follows genetic mechanisms of antibody diversity.

Original interface screenshots are reproduced in their original colors without cropping; the adjacent notes are analytic annotations. Each breadcrumb makes the local continuation visible, while the route context identifies its narrower explanatory focus. Neither displayed page provides visible source attribution for its specialist claims. The screenshots document generated content and are not verified immunology illustrations.

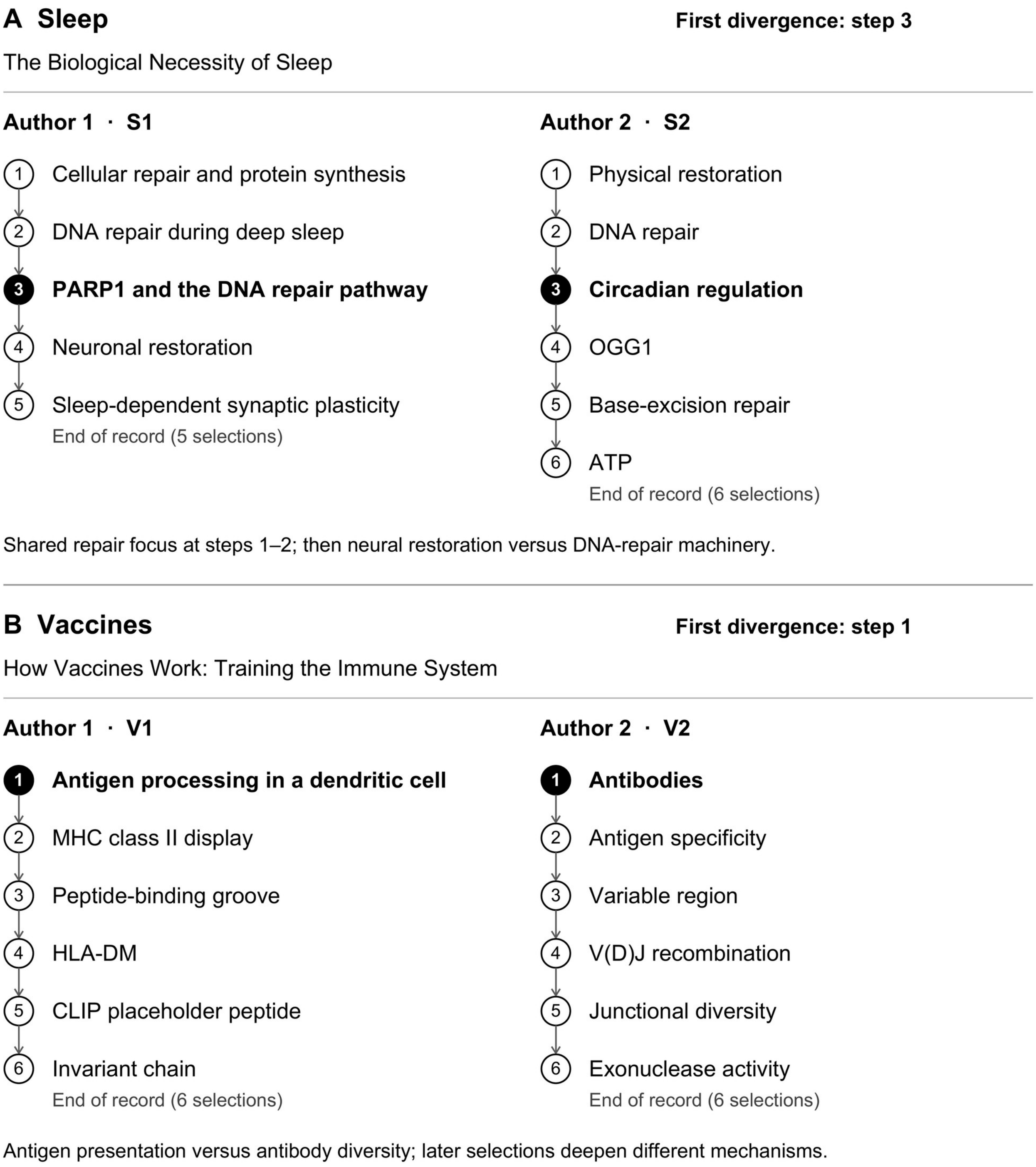

Figure 2. Recorded navigation routes (A–B). Read downward within each author’s lane and compare the same step across lanes. Numbered nodes identify selections; filled nodes and bold labels mark the first divergence in explanatory focus. Each pair shares the prompt printed above it, although generated entry pages and branch menus could differ. Only selected paths are shown. Route IDs and step numbers link to Table 1.

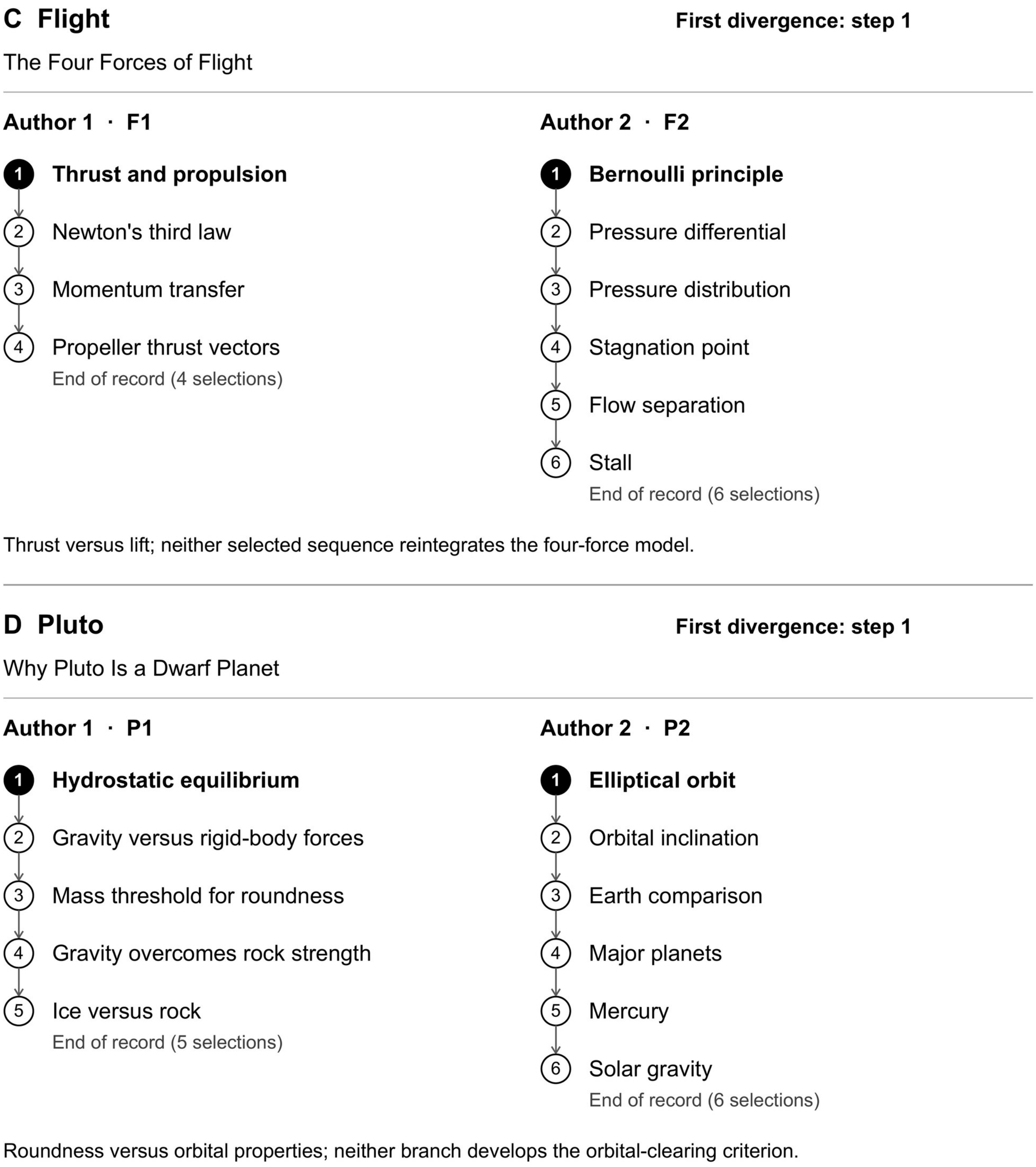

Figure 2 (continued). Flight and Pluto routes (C–D), using the same notation. The end of each record is shown at its observed depth; empty space below an endpoint represents no further recorded selections. The diagram describes route structure, not conceptual strength, comprehension, or transfer.

## Appendix A. Analytic Distinctions and Evidence Locations

The definitions below articulate the qualitative distinctions used to read the documented routes. A transition can narrow to a submechanism while remaining locally coherent and relevant to part of its starting question.

**Table A1. Units and interpretive decision boundaries.**

| Distinction | Unit | Interpretive boundary |
|---|---|---|
| Convergence and divergence | Paired routes | Compare the mechanism or explanatory level selected from the shared prompt. Similar wording alone does not establish convergence. |
| Local coherence | Adjacent pages | Identify the relation connecting consecutive pages, such as structure to function, process to component, or comparison. Topical similarity alone is insufficient. |
| Narrowing | Successive selections | Identify a shift from a broader topic to one component, mechanism, or explanatory scale. Narrowing can remain relevant to the prompt. |
| Repetition | Adjacent pages | Identify overlapping explanatory content and assess what the later page adds. Related titles alone do not establish repetition. |
| Global alignment and drift | Route with its prompt | Ask which part of the original question the accumulated explanation addresses. Separate partial coverage from a sequence that no longer advances that question. |
| Visible claim provenance | Page or claim | Locate the displayed attribution supporting a claim. Absence of a visible source is a provenance observation, separate from a judgment of scientific accuracy. |

**Table A2. Source key for route and step identifiers.**

| Prompt | Author 1 reconstruction | Author 2 structured log |
|---|---|---|
| Sleep | S1: Author 1 – Sleep Walkthrough | S2: W03 |
| Vaccines | V1: Author 1 – Vaccines Walkthrough | V2: W05 |
| Flight | F1: Author 1 – Flight Walkthrough | F2: W06 |
| Pluto | P1: Author 1 – Pluto Walkthrough | P2: W04 |

Each record contains an initial screenshot and numbered click entries. Author 1’s reconstructions describe the screenshot evidence; Author 2’s logs also include contemporaneous selection rationales. Author 2’s post-walkthrough judgments are recorded in Tasks 4: Knowledge Maps. Table 1 brings representative transitions together with route-level interpretations.